\documentclass[aps,prd,preprint,groupedaddress,amsfonts,amssymb,amsmath,showpacs]{revtex4}

\usepackage{amsmath,amssymb}

\usepackage{graphicx}

\usepackage{psfrag}

\usepackage[usenames]{color}

\newcommand{\km}{k_{\rm M}}
\newcommand{\kmi}{k_{\rm min}}
\newcommand{\kx}{k_{\rm max}}
\newcommand{\gr}{{\rm g}}
\newcommand{\R}{\mathcal{R}}

\newcommand{\Pe}{\mathcal{P}}

\newcommand{\zetar}{\zeta_{\R}}
\newcommand{\Sigmar}{\Sigma_{\R}}
\newcommand{\e}{\times 10^}
\newcommand{\ten}{10^}
\newcommand{\mpbh}{M_{\rm PBH}}
\definecolor{myred}{rgb}{0.85,0.08,0}
\definecolor{mydb}{rgb}{0,0.08,0.8}
\long\def\symbolfootnote[#1]#2{\begingroup%
\def\thefootnote{\fnsymbol{footnote}}\footnote[#1]{#2}\endgroup}

\newcommand{\beq}{\begin{equation}}
\newcommand{\eeq}{\end{equation}}
\newcommand{\bea}{\begin{eqnarray}}
\newcommand{\eea}{\end{eqnarray}}
\def\bi{\begin{itemize}}
\def\ei{\end{itemize}}

\def\bx{{\bf x}}

\def\mP{m_{\rm Pl}}
\def\fnl{f_{\rm NL}}

\newcommand{\gsim}{\ \raise.3ex\hbox{$>$\kern-.75em\lower1ex\hbox{$\sim$}} \ }
\newcommand{\lsim}{\
  \raise.3ex\hbox{$<$\kern-.75em\lower1ex\hbox{$\sim$}} \ }
\begin{document}

\title{The effect of non-Gaussian curvature perturbations on the formation of primordial black holes}
\author{J.~C.~Hidalgo$^{1}$} \email[Email
  address:]{c.hidalgo@qmul.ac.uk}
\affiliation{$^1$ Astronomy Unit, School of Mathematical Sciences,
  Queen Mary, University of London, Mile End  Road, London E1 4NS,
  United Kingdom }

\date{\today}

\begin{abstract}

This paper explores the consequences of non-Gaussian cosmological
perturbations for the formation of primordial black holes (PBHs).
A non-Gaussian probability distribution function (PDF) of
curvature perturbations  is presented with an explicit
contribution from the three-point correlation function to linear order. The
consequences of this non-Gaussian PDF for the large perturbations
that form PBHs are then studied. Using the observational limits
for the non-Gaussian parameter $f_{NL}$, new bounds to the mean
amplitude of curvature perturbations are derived in the range of scales
relevant for PBH formation.
\end{abstract}


\maketitle

\section{Introduction}\label{sec1}

The non-linear theory of perturbations is an important part of the study
of the early universe because only at this level of precision one can
distinguish specific models of inflation and even rule out some of
them through observations of the correspondent non-Gaussian statistics. 
The possibility of observing non-Gaussian correlations in the
temperature fluctuations of the cosmic microwave background radiation 
(CMB) \cite{spergel06} and in the count of large scale structure
(LSS)\cite{matarrese00,bernardeau01,kang07}
has fostered the study of non-linear perturbations for distinct
models of the origin of structure
(see e.g. Refs. \cite{lyth05c,chen06,seery06a}, and  see
Ref. \cite[][]{bartolo04} for an extended review of both
observational and theoretical aspects of non-Gaussianity,).

The properties of the early universe imprinted in the observed
perturbations are better studied when we make use of the curvature
perturbation $\R(t,\bx)$ on comoving
hypersurfaces 
\footnotemark.\footnotetext{\footnotesize In
  the longitudinal gauge, the scalar isotropic curvature perturbation or
  Bardeen potential $\Phi$ modifies the Friedmann background
  metric to the form: 
  \begin{align*}
    ds^2 =  -\left[ 1 + 2 \Phi(t,\bx) \right] dt^2 + a(t)^2
    \left[ 1 - 2 \Phi(t,\bx) \right]   d{\bf x}^2.
  \end{align*}
  \noindent where $t$ is the cosmic time,
  $a(t)$ is the scale factor and $d\bx$ is the comoving spatial
  part. 
  The comoving curvature perturbation $\R(t,\bx)$ is defined in
  the comoving gauge from the metric
  \begin{align*}
    ds^2 =  - N^2( t , \bx ) dt^2 + a(t)^2 e^{ 2 \R(t,\bx) }
    \delta_{ i j }  \left( dx^i + N^i ( t , \bx ) dt \right)
    \left( dx^j + N^j ( t , \bx ) dt \right),
  \end{align*}
  \noindent where $\delta_{ij}$ is the Kronecker delta,
  and $N$ and $N^i$ are functions algebraically determined
  from the symmetries of the perturbed equations of motion.}
This quantity is defined in a
gauge-invariant way
and, when sourced by adiabatic matter
perturbations, the growing perturbation mode remains constant
on scales larger than the cosmological horizon $H^{-1}$, where $H =
d\ln{(a)} / d t $ is the Hubble parameter \cite{wands00,lyth05}.

A fundamental observable of cosmological perturbations
is the mean amplitude. The classical amplitude of metric perturbations
$\R_{\rm cl}$ is derived by solving the perturbed Einstein
equations to linear order. Statistically this classical amplitude
is written in terms of the two point correlation function as \cite{kolb90},
\begin{align}
  \langle \R_{\rm G}({\bf k}_1)\R_{\rm G}({\bf k}_2)\rangle =
  (2\pi)^3  \delta({\bf k}_1 + {\bf k}_2) |\R_{\rm cl}(k)|^2,
  \label{eqn1.33}
\end{align}

\noindent where $\R_{\rm G}({\bf k})$ are Gaussian
perturbations in Fourier space, and ${\bf k}_1$ and ${\bf k}_2$
are vectorial Fourier momenta. The two-point
correlator defines also  the dimensionless
power-spectrum $\Pe(k)$ as
\begin{align}
  \label{eqn1.4}
  \langle \R_{\rm G}({\bf k}_1)\R_{\rm G}({\bf k}_2)\rangle =
  (2\pi)^3  \delta({\bf k}_1 + {\bf k}_2)\frac{2\pi^2}{k_1^3} \Pe(k_1).
\end{align}

The power-spectrum encodes important information of
the underlying cosmological model. For example, for perturbations
generated from a single inflationary field $\phi$ with a potential 
$V$ dominating the cosmological dynamics, $\R_{\rm G}(k)$ is a
random field of perturbations about the quasi-de Sitter
background, and the power-spectrum is given by \cite{stewart93}
\begin{align}\label{eqn1.5}
  \Pe(k) = \frac{H_*^4}{(2\pi)^2 \dot{\phi}_*^2 \mP^2}
  \approx \frac{V_*^3}{(d V / d\phi)_*^2\mP^2},
\end{align}

\noindent where $\mP = 8\pi G$ is the Planck mass and where an
asterisk denotes evaluation at $t_*$,
the time when the relevant perturbation mode exits the
cosmological horizon, $k = a_* H_* = a(t_*) H(t_*)$.

The tilt of the power-spectrum is
parametrised with a second observable, the spectral index
which is defined as
\begin{align}\label{eq1.55}
  n_s = \frac{d}{d \ln{k}}\ln{\Pe(k)}.
\end{align}

\noindent With this definition, $n_s < 0 $ means that the
power-spectrum is larger for large scales, in such case we
have a red spectrum. Equivalently $n_s >0$ corresponds to a
larger power for small scales and this is called a blue
spectrum.

The power-spectrum and the tilt are
derived directly from linear perturbations. Assuming linearity
for the perturbations of the CMB, it is possible to determine
with great accuracy the numerical values of the
power-spectrum and its tilt on scales larger than the horizon at the
time of last scattering. At such scales, the most recent
observations of the CMB and the galaxy count give $\Pe = 2.4\times
10^{-9}$, $n_s = - 0.05 \pm 0.01$ \cite{spergel06}. Any successful
theory of structure formation must meet these values at the relevant
scales. Higher order correlations of the perturbation field $\R$ offer
an exciting way to distinguish between these cosmological models.

The deviations from Gaussianity are described to lowest order by the
non-linear parameter $\fnl$. Mathematically, this parameter
appears in the expansion \cite{lyth05a} 
\begin{align}
  \label{eqn1.3}
  \R(k) = \R_{\rm G}(k) - \frac{3}{5} \fnl (\R_{\rm G}\star\R_{\rm G}(k) -
  \langle \R_{\rm G}^2 \rangle),
\end{align}

\noindent where a star denotes a convolution of two copies of the field. 
We use this definition throughout the paper. At the present time,
the precision of the observations can
only provide limits for $\fnl$. The WMAP
satellite gives the constraints $-54 < \fnl < 114$
at the $95\%$ confidence level \cite{spergel06}.

When one computes second-order perturbations to
the Einstein equations, higher order correlations
become non-vanishing
(see e.g. Refs. \cite[][]{bartolo03,rigopoulos04}).
Alternatively such correlator can be obtained from the third-order
quantum perturbations to 
the Einstein-Hilbert action
(Pioneering works using this method are
Refs. \cite{maldacena02,seery05a,seery05b}).
The three-point correlation function in Fourier space at the tree
level can be directly computed from the definitions in
Eqs. \eqref{eqn1.33}\eqref{eqn1.4} and \eqref{eqn1.3} to find
\cite{lyth05a} 
\begin{align}
  \langle \R(k_1)\R(k_2)\R(k_3)\rangle
  =&   - ( 2 \pi )^3  \delta \left( \sum_i {\bf k}_i  \right)  4  \pi^4
  \frac{6}{5}  \fnl  \left[ \frac{\Pe(k_1) \Pe(k_2)}{k_1^3 k_2^3}
    + 2 \mbox{ perm} \right]. \label{eqn1.6}
\end{align}

Non-linear correlations have
been explored not only in the CMB anisotropies
\cite{spergel06, komatsu01}, but also in counts of galaxies
and clusters
\cite{matarrese00,scoccimarro03,hikage06,kang07}.
The latter deserve special mention.
The distribution of galaxies and clusters account for
density perturbations that formed such structures in the matter-dominated
era. The statistics of these objects gives us extra
information about the primordial non-Gaussianities in the
universe. However, in studying the formation of galaxies and
clusters from the seed perturbations, one must be careful in
the treatment of intermediate stages because the
growth of perturbations after horizon entry is intrinsically
non-linear. This may blur the primordial non-Gaussian statistical effects
(For a review, see Ref. \cite[][]{bernardeau01}).

In this paper we study non-Gaussianity for a range of wavelengths
not covered by the observations mentioned above by looking at the
formation of Primordial Black Holes (PBHs). PBHs were
formed during the radiation era of our universe from large
amplitude perturbations. The importance of non-Gaussianity in this
context is that a small deviation from linearity would change the
probability of formation significantly. The effects of
non-Gaussian perturbations on PBHs have been studied for specific
models \cite{bullock96,ivanov98, avelino05} but a
precise quantification of the non-Gaussian effects is still required
and it is only now, with a much better understanding of the effects
of higher order perturbations, that we are able to describe the general
effects on PBHs. The discussion of non-Gaussian effects on PBH
formation turns crucial in the light of recent works
(e.g. \cite{siri06}) which claim that only through exotic extensions
of the canonical slow-roll inflationary potentials can produce considerable
amounts of PBHs. Here we explore whether non-Gaussian perturbations
could ease this dificulty.

To look at the effects of non-Gaussianity we make use of a
non-Gaussian PDF for curvature perturbations with a direct dependence
on the non-linear parameter $\fnl$ as derived in \cite{sh06}. We adapt
this PDF to calculate probabilities for pertrbations relevant to PBH
formation. We also explore the amount of  non-Gaussianity that is
allowed from the bounds to the total mass fraction of PBHs. 
In principle, the PBH abundance could be used to probe
non-Gaussianity on scales not reached through CMB or LSS
observations. We find that even though the possible limits on
$\fnl$ in the relevant mass regime are weak, its current allowed
values in the CMB scale can be used to generate new bounds on the
mean amplitude of curvature perturbations $\R$ for
the wavelength range $ 10^{-16} <  \lambda / Mpc  < 1 $.

This paper is organised as follows. In Section \ref{sec2} we construct the
appropriate non-Gaussian probability distribution function
(PDF) to calculate the probability of non-linear perturbations
forming PBHs. In Section \ref{sec3} we characterise the effects of the
non-Gaussian PDF on the probability of PBH formation. We recover
the results presented in previous studies \cite{ivanov98,
  bullock96} and reconcile the discrepancy in their conclusions. In
Section \ref{sec4}, we make use of the Press-Schechter formalism to derive
the abundance of PBHs including non-Gaussian effects. We show how
the constraint on the amplitude of perturbations is modified by
the introduction of a non-Gaussian mass function and report on new
bounds on the variance of curvature perturbations. We conclude in
Section \ref{sec5} with a summary of results.

\section{The non-Gaussian PDF}\label{sec2}

In linear perturbation theory one makes use of the central
limit theorem to construct the probability distribution function
(PDF). To first order, the perturbation modes are independent of
each other. If we define the field of perturbations $\R$ with 
zero spatial average, then the central limit theorem indicates
that the PDF of $\R$ is a normal distribution dependent
on a single parameter, the variance $\langle\R^2\rangle$,
\beq
\mathbb{P}_{\rm G}(\R) \approx \frac{1}{\sqrt{\langle\R^2\rangle}}
\exp\left(-\frac{\R^2}{2\langle\R^2\rangle}\right).
\label{eqn2.0}
\eeq

An immediate effect of including the non-linearity of
the perturbations is the interaction of distinct
perturbation modes. Indeed, as shown in Eq. \eqref{eqn1.3},
the parametrisation of higher order
contributions to $\R$ is described in terms of convolutions in
Fourier space. The requirements of the central limit theorem are
not met, so it can no longer be used to construct the PDF.

In a recent paper a first-order correction to the Gaussian PDF
was explored, and a new PDF was derived, which includes the linear
order contribution from the 3-point function. In the following we describe the
elements of such derivation.

As is customary in the treatment of perturbations, we
first smooth the field over a given mass scale $\km$
with a window function $W_{\rm M}(k)$,
\begin{align}
  {\bar \R}({\bf x}) = \int \,\frac{d{\bf k}}{(2\pi)^3}
  W_{\rm M}(k) \R({\bf k}) e^{i({\bf k\cdot x})}. \label{eqn2.1}
\end{align}

\noindent Here we use a truncated Gaussian window function:
\begin{align}
  \label{eqn2.11}
  W_{\rm M}(k) = \Theta(\kx - k)\exp{\left(-\frac{k^2}{\km^2}\right)},
\end{align}

\noindent where $\Theta$ is the Heaviside function and
the fiducial scale $\kx$ is  introduced to
avoid ultraviolet divergences. The amplitude of the
perturbation is parametrised by the central value of the
configuration ${\bar \R}(\mathbf{x} = 0) = \zeta_{\R}$. This
parametrisation is particularly useful when we pick the relevant
perturbations for the formation of PBHs \cite{shibata99}. The
non-Gaussian probability distribution function for a perturbation with
central amplitude $\zeta_{\R}$ is \cite{sh06}
\begin{equation}
  {\mathbb P}_{\rm NG}(\zetar)    =
  \mathbb{P}_{\rm G}(\zetar) \left[1 + \left(\frac{\zetar^3}{\Sigmar^3}
    -  \frac{ 3\,
      \zetar}{\Sigmar}\right)\frac{\mathcal{J}}{\Sigmar^3}
    \right],\label{eqn2.2}
\end{equation}

\noindent where $\Sigmar^2$ is the variance of the smoothed field,
related to the power-spectrum by
\begin{equation}
  \Sigmar^2(M) = \int \, \frac{dk}{k} W_{\rm M}^2(k) \Pe(k),
  \label{eqn2.22}
\end{equation}

\noindent and the factor $\mathcal{J}$ encodes the non-Gaussian
contribution to the PDF:
\begin{align}
  \mathcal{J}  &= \frac{1}{6  }\int
  \,\frac{d{\bf k}_1 \,d{\bf k}_2 \,d{\bf k}_3}{(2\pi)^9}
  W_{\rm M}(k_1)W_{\rm M}(k_2)W_{\rm M}(k_3) \langle\R({\bf k}_1)
  \R({\bf k}_2)\R({\bf k}_3)\rangle,
  \label{eqn2.4} \\
  &=   - \frac{1}{5}\int \,\frac{\,d{\bf k}_1
    \,d{\bf k}_2 \,d{\bf k}_3    }{(4  \pi)^2
    \prod_i     {W}_{\rm M}^{-1}({ k}_i)}
  \delta\left(\sum_i {\bf k}_i\right) \; \fnl
  \left[\frac{\Pe(k_1)\Pe(k_2)}{k_1k_2} + 2
    \mbox{ perm.}\right], \label{eqn2.3}      
\end{align}

\noindent where the last equation is valid at tree level in the
expansion of $\langle\R\R\R\rangle$. This limitation is justified
as long as the loop contributions to the three-point function,
generated from the convolution of $\R$-modes, are
sub-dominant. Such is the case for $\fnl \leqslant 1/ \Pe(k)$
for all values of $k$. As will be seen in section \ref{sec4}, the tested
$\fnl$ does not exceed such values.

Further details of the derivation of the PDF in Eq. \eqref{eqn2.2} can be
found in Ref. \cite{sh06}. Here is sufficient to say that the
time-dependence of this probability is eliminated when the averaging
scale is $\km \leq a(t)H(t)$ because the growing mode of the 
perturbation $\R$ is constant on superhorizon scales \cite{wands00,lyth05}.

We now proceed to integrate the non-Gaussian factor in
Eq. \eqref{eqn2.3} over the amplitudes and scales 
relevant to PBHs. The expression in Eq. \eqref{eqn2.3} is
integrated within the limits  $\kmi$ and $\kx$ defined conveniently
to cover the relevant perturbation modes for PBH formation. These
objects are formed long before the matter-radiation equality, so
in the large-box (small wavenumber) limit of the integral
\eqref{eqn2.3} we can choose $\kmi = H_0$, the Hubble horizon
today. This is a reasonable lower limit for integrating 
perturbations relevant for PBH formation \cite{lyth91}. At the other end
of the spectrum, the smallest PBHs have  the size  of the Hubble
horizon  at the end  of inflation. We therefore use $\kx = a(t_{\rm END})
H_{\rm END}  $, the comoving horizon at the end of inflation, as a
suitable upper limit.  It is important to mention that, even though the
integral in Eq. \eqref{eqn2.3} should add all
$k$-modes, finite limits are imposed to avoid logarithmic
divergences. Moreover, due to the window function $W_{\rm M}(k)$,
the dominant part of the integral is independent of the choice of
integration limits as long as they remain finite. 

We solve integral \eqref{eqn2.3} for the limit of equilateral
configurations of the three-point correlator, that is,
correlations for which  $k_1 = k_2  = k_3 $. This is not merely a
computational simplification. In the integral, each perturbation
mode has a filter factor $W_{\rm M}(k)$ which, upon integration, picks
dominant contributions from the smoothing scale $\km$
common to all perturbation modes \footnotemark. We take this
argument as an ansatz, in which case $\mathcal{J}$ can be written
in the suggestive way:
\footnotetext { \footnotesize This is an
  important restriction to the kind of non-Gaussianities that can be
  tested by PBHs \cite{babich04}. If a strong non-Gaussian
  signature is encountered exclusively for isosceles triangulations, this
  cannot affect the formation of PBHs.}
\begin{equation}
  \mathcal{J}  =   -  \frac{1}{8}  \int_{\kmi}^{\kx}\,\frac{dk}{k}
  \left[W_{\rm M}(k)\Pe(k)\right]^2 \left( \frac{6}{5}\fnl\right).
  \label{eqn2.6}
\end{equation}

In the following sections  we compute $\mathcal{J}$ numerically
for inflationary perturbations generated in a single field
slow-roll inflationary epoch, and for the case of constant
$\fnl$. This will be used to test the effects of
non-Gaussianity on PBHs.

\section{Non-Gaussian modifications to the probability of PBH
  formation} \label{sec3}

The simplest models of structure formation within the inflationary
paradigm are those where a single scalar field drives
an accelerated expansion of the spacetime and its quantum
fluctuations give birth to the observed structure in
subsequent stages of the universe (For an extended review of
the inflationary paradigm, see Ref. \cite[][]{liddle00}). Here we
study the curvature perturbations generated at the time of single
scalar-field inflation including contributions from non-linear
perturbations.

The effects of non-Gaussianity on PBHs have been explored in the past
but with inconclusive results.
Bullock and Primack \cite{bullock96} studied the formation of PBHs numerically
for perturbations with blue spectra ($n_s > 0 $) and non-Gaussian
contributions.
The motivation for this was that any inflationary
model with a constant tilt and consistent with the normalisation of
perturbations at the CMB scale, must have a blue spectrum to produce a
significant number of PBHs \cite{lidsey94,green97}. Their
analysis is based on the stochastic generation of perturbations on
superhorizon scales, together with a Langevin equation for
computing the PDF. For all the cases tested, the non-Gaussian PDF
is skewed towards small fluctuations, so PBH production, which
integrates the high amplitude tail, is suppressed with respect
to the Gaussian case. An example of the kind of potentials
studied in Ref. \cite{bullock96} is
\begin{equation}\label{eqn3.4}
  V_1(\phi) = V_0\left\{\begin{array}{rcl}
  1 + \arctan\left(\frac{\phi}{\mP}\right),
  & \mbox{for} & \phi > 0, \\
  1 + (4\mbox{x}10^{33})\left(\frac{\phi}{\mP}\right)^{21},
  & \mbox{for} & \phi < 0.
  \end{array}\right.
\end{equation}

\noindent where $V_0$ is the amplitude of the potential at $\phi =
0$.

Another way of generating large perturbations in the inflationary
scenario is to consider localised features in the potential dominating the
dynamics. As one can see from Eq. \eqref{eqn1.5}, an abrupt
change in the potential would generate a spike in the spectrum of
perturbations. The effects of non-Gaussianity for an inflationary
model producing features in an otherwise red spectrum ($n_s < 0$)
were explored by \citet{ivanov98} using the toy model
\begin{equation} \label{eqn3.5}
  V_2(\phi) = \left\{\begin{array}{rcl}
  \lambda \frac{\phi^4}{4}\quad & \mbox{for} & \phi < \phi_1, \\
  A(\phi_2 - \phi)  +\lambda \frac{\phi_2^4}{4} & \mbox{for} &
  \phi_2 > \phi >\phi_1, \\
  \tilde{\lambda} \frac{\phi^4}{4} \quad& \mbox{for} & \phi > \phi_2.
  \end{array}\right.
\end{equation}

\noindent where $\lambda$ and $\tilde{\lambda}$ are coupling constants.
Through a stochastic computation of the PDF, it was found that
large amplitude perturbations were more abundant for a non-Gaussian
PDF than for a Gaussian one.

To understand this difference and  generalise the
effects of non-Gaussianity, we look at the fractional difference
of the Gaussian and non-Gaussian PDFs:
\begin{equation}
  \label{eqn3.7}
  \frac{\mathbb{P}_{\rm NG} -\mathbb{P}_{\rm G}}{\mathbb{P}_{\rm G}} =
  \left[\left( \frac{\zetar^3}{\Sigmar^3} -
    3 \frac{\zetar}{\Sigmar}\right)\frac{\mathcal{J}}{\Sigmar^3}\right].
\end{equation}

\noindent Note that both Refs. \cite{bullock96} and \cite{ivanov98} use
perturbations generated in a piecewise slow-roll inflationary
potential for which inflation is controlled by keeping the slow-roll
parameters $\epsilon \equiv 1/2(\mP V'/ V)^2$ and
$\eta \equiv \mP^2 (V''/V)$ smaller than one. Here we make use of the
slow-roll approximation to explore the qualitative effects of
Eq. \eqref{eqn3.7}. To linear order,
there is a straightforward expression for the spectral index in
terms of these parameters \cite{stewart93},
\begin{align}
  \label{eqn3.71}
  n_s   =  2(\eta - 3 \epsilon).
\end{align}

\noindent On the other hand, carrying a first order expansion on slow-roll
parameters, Maldacena, provides an expression for the
non-linear factor $\fnl$ in terms of the slow-roll parameters
 \cite{maldacena02}, 
\begin{align}
  \label{eqn3.8}
  \fnl = &\frac{5}{12}\left( n_s + \mathcal{F}(k) n_t\right) =
  \frac{5}{6}\left( \eta - 3 \epsilon + 2 \mathcal{F}(k) \epsilon \right),
\end{align}

\noindent where $n_t = 2\epsilon$ is the scalar-tensor perturbation
tilt and $\mathcal{F}(k)$ is a number depending on the triangulation
used. For the case of equilateral configurations, when $\mathcal{F}
=  5/6$,
\begin{align}
  \fnl = & \frac{5}{6}\left(\eta - \frac{4}{3}\epsilon \right)_{\rm eq}.
\end{align}

\begin{figure}[hb]
  \begin{center}
    \psfrag{A}{\Large$\frac{\mathbb{P}_{\rm NG}(\zeta) - \mathbb{P}_{\rm
	  G}(\zeta)}{\mathbb{P}_{\rm G}(\zeta)}$}
    \psfrag{B}{\Large$\zeta$}
    \includegraphics[totalheight=0.35\textheight]{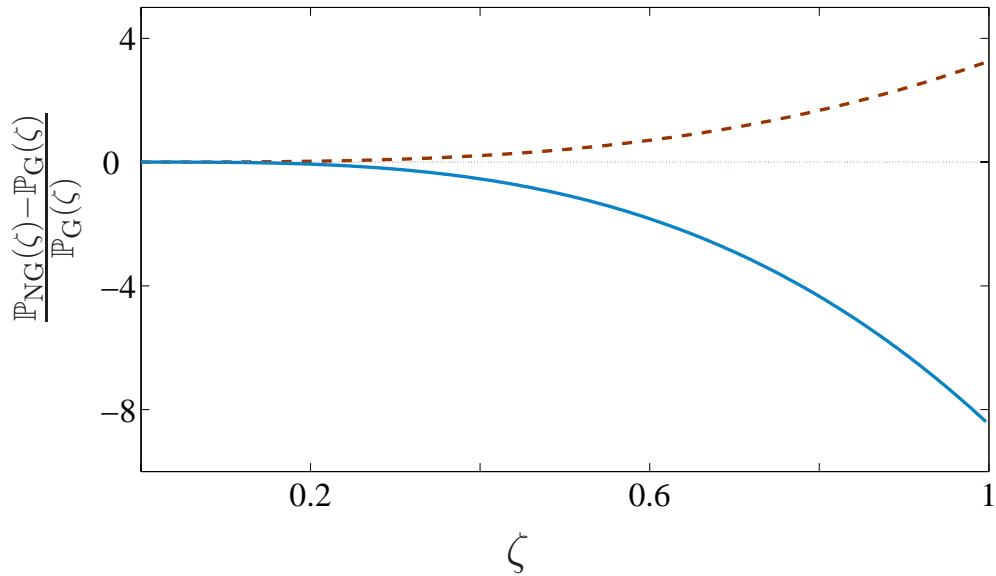}
    \\
    \caption{{The fractional departure from Gaussianity is
	plotted for two types of non-Gaussian distributions $P_{NG}$
	\eqref{eqn2.2}. For the potential in Eq. \eqref{eqn3.4}, $\fnl >
	0$. The potential of Eq. \eqref{eqn3.5} gives $\fnl < 0$ and its
	correspondent PDF is shown by a dashed line. }}\label{fig1}
  \end{center}
\end{figure}

\noindent We use this last result to evaluate the integral 
\eqref{eqn2.6}. The non-Gaussian effect on the PDF is illustrated in
Fig.\ref{fig1} for the potentials given by Eqs. \eqref{eqn3.4} and
\eqref{eqn3.5} in terms of the fractional difference
\eqref{eqn3.7}. This last factor represents the skewness of the
non-Gaussian PDF. Consequently the sign of $\fnl$ is what determines
the enhancement or suppression of the probability
for large amplitudes $\zetar$ in our  non-Gaussian PDF. Indeed, the
non-Gaussian contribution encoded in the factor $\mathcal{J}$ is the
sum of the $\fnl$ value over all scales relevant for PBH
formation. For the two cases illustrated, the scalar tilt $n_s$ dominates
over the tensor tilt $n_t$ in a way that the sign of $\fnl$
incidentally coincides with that of $n_s$.

\section{Constraints on non-Gaussian perturbations in the PBH
  range}\label{sec4}

Primordial Black Holes are objects that collapsed from
large-amplitude perturbations at times previous to photon
decoupling \cite{carr75}. The energy density of PBHs formed during
inflation is diluted by the superluminous expansion. In consequence, a
significant production of PBHs can only take place after
inflation. Depending on the model of inflation,
PBHs can cover a wide range of masses $ M_{\rm END}\sim 10^{-48}
M_{\odot} \leq \mpbh \leq 10 M_{\odot}$ at scales much smaller than
galaxies $M \sim 10^{9} M_{\odot}$ or clusters $M \sim 10^{13}
M_{\odot}$. We will call this the PBH mass range.
The lack of direct observations of PBHs limits their
cosmological abundance. Assuming they are nonetheless present,
there are three ways in which PBHs can affect the evolution of our
universe and converselly, we can impose constraints to the
abundance of such objects. First, the 
current density of PBHs cannot exceed the amount of dark matter
density, i.e., $\Omega_{\rm {PBH}}(M\geq 10^{15} \gr) \leq
\Omega_{\rm DM} = 0.47 $ (with the WMAP-III value taken at
$3\sigma$ confidence level \cite{spergel06}). Second, the Hawking
radiation \cite{hawking74} from PBHs, can also generate the radiation
observed at various wavelengths in our universe \cite{carr76}. The
lifetime $t_{\rm evap}$ for PBHs is a function of its mass 
\begin{align}
  t_{\rm evap} = 1.2\e{-44}\left(\frac{M}{\mP}\right)^3
  \text{sec}.\label{eqn4.1}
\end{align}

\noindent From this relation we immediately infer that PBHs of mass
$M_{\rm evap}= 5\times 10^{14}\gr$ 
are evaporating today and the observation of the gamma-ray
background constrains their present density parameter to $\Omega_{\rm
  PBH}(M_{\rm evap}) \lesssim 5 \times 10^{-8}$ \cite{page76,carr76,
  macgibbon91,macgibbon99}. This is the tightest constraint on the
abundance of PBHs. A third cathegory of constraints is relevant for
lighter PBHs. Black 
holes with mass $M < M_{\rm evap}$ have already evaporated and the
decay products should not spoil the well understood chemical history
of our universe (see e.g. Refs.
\cite{miyama78, novikov79}).

To calculate the PBH mass fraction we make use of a standard
Press-Schechter formalism \cite{press74}. This
formula integrates the probability of PBH formation over the
relevant matter perturbation amplitudes \cite{carr75}. This is
interpreted as the mass fraction of such PBHs at the time of formation,
\begin{align}
  \beta_{\rm PBH}(\geq M) = 2 \int_{\delta_{\rm th}}^{\infty}
  \mathbb{P}(\delta(M))\,d \delta(M), \label{eqn4.2}\\
  \approx \frac{\sigma_{\delta}(M)}{\delta_{\rm th}}
  \text{exp}\left[-\frac{\delta_{\rm th}^2}{2 \sigma_{\delta}^2(M)}\right].
  \label{eqn4.22}
\end{align}

\noindent Here $\delta = \delta \rho / \rho$ is the
matter density perturbation $\sigma^2_{\delta}$ is the corresponding
variance and $\delta_{\rm th}$ is the
threshold amplitude of the perturbation necessary to form a PBH.
By integrating over a smoothed perturbation, this integral is
equivalent to the mass fraction of PBHs of mass $M \geq
\gamma^{3 / 2} M_{H} \approx \gamma^{3 / 2} \km / (2\pi)$
\cite{carr75},
where $\gamma$ is the sound-speed squared at the time
formation. Note that the approximation \eqref{eqn4.22}
is valid only for a Gaussian PDF.

The integral \eqref{eqn4.2} establishes a direct relation between the
mass fraction of PBHs and the variance of perturbations. The set of
observational constraints on the abundance of PBHs is listed
in Table I and has been used to bound the mean amplitude of
$\delta$ defined by for distinct cosmologies
\cite{lidsey94,green97,clancy03,sendouda06}. The
Press-Schechter formula has also been tested against other
methods such as peaks theory \cite{liddle04}.

The threshold value $\delta_{\rm th}$ used in Eq. \eqref{eqn4.2}
has changed with the improvement of gravitational collapse studies
\cite{carr75, niemeyer97, shibata99, hawke02, musco04}. Here we use
the value $\delta_{\rm th}=0.3$ for convenience
\footnotemark.
\footnotetext{\footnotesize Recent studies suggest
  that this value is dependent on the profile of the curvature
  perturbation \cite{musco07}. Furthermore some early universe
  models may not generate the profiles required for PBH formation.
  This is a crucial aspect of PBH formation currently under investigation
  \cite{hidalgo07}} The corresponding threshold value of the
curvature perturbation can be deduced from the relation
\cite{liddle00}
\begin{align}
  \delta_{k}(t) = \frac{2(1+\gamma)}{5 + 3\gamma}
  \left(\frac{k}{aH}\right)^2 \R_k,\label{eqn4.3}
\end{align}

\noindent which at horizon crossing during the radiation
dominated era gives $\R_{\rm th} = 0.7$. Note that the high
amplitude of the perturbations relevant
to PBH formation necessarily require a non-linear treatment of
their statistics. This is the major motivation for our analysis.

We adopt the Press-Schechter formula derive the non-Gaussian abundance of
PBHs. The use of the Press-Schechter integral on distributions of curvature
perturbations is not new. In Ref. \cite{zaballa06} the Press-Schechter
formula is used to integrate curvature perturbations which never
exit the cosmological horizon. We apply the integral formula
in Eq. \eqref{eqn4.2} to the non-Gaussian probability
distribution \eqref{eqn2.2}. The
result of the integral is the sum of incomplete Gamma functions
$\Gamma_{\rm inc}$ and an exponential:
\begin{align}
  \begin{split}
    \beta(M) = 
    \frac{1}{\sqrt{4\pi}}& \Gamma_{\rm inc}\left(1/2,
    \frac{\zeta_{\rm th}^2}{2\Sigmar^2(M)}\right) - \\
    &  \frac{1}{\sqrt{2\pi}}\frac{\mathcal{J}}{\Sigmar^3(M)}
    \left[2\,\Gamma_{\rm inc}\left(2,\frac{\zeta_{\rm
	  th}^2}{2\Sigmar^2(M)} \right) - {3} \exp{\left(- \frac{\zeta^2}{2
	  \Sigmar^2(M)}\right)}  \right].\label{eqn4.31}
  \end{split}
\end{align}

\noindent The Taylor series expansion of these functions
around the limit $ \Sigmar / \zeta_{\R} = 0$ gives
\begin{align}
  \beta(M) \approx \frac{\Sigmar(M)}{\zeta_{\rm
      th}\sqrt{2\pi}}
  \exp \left[-\frac{1}{2}\frac{\zeta_{\rm th}^2}{\Sigmar^2(M)}\right] \times
  \left\{ 1 -  2\left(\frac{ \Sigmar}{\zeta_{\rm th}}\right)^2  +
  \frac{\mathcal{J}}{ \Sigmar^3} \left[ \left(\frac{ \Sigmar}{\zeta_{\rm
      th}}\right)^{-2} - 1 \right] \right\}. \label{eqn4.4}  
\end{align}

For the mass fraction shown in Eq. \eqref{eqn4.4}, the
observational limits of Table I could in principle constrain the
values of the variance $\Sigmar^2$ and of $\fnl$. However,
when we normalise the amplitude of perturbations to the observations at
CMB scales, the limiting values for $\fnl$ are too large, of
order $\fnl \approx 10^4$,
and thus inconsistent with perturbation theory at the level of the
expansions performed in this paper. In fact, the expansion in Eq.
\eqref{eqn1.3} shows that when 
\begin{align}
  \left|\fnl \right| \geq \frac{1}{\R^2} \approx \frac{1}{\Pe(k)}
  \approx 203.2,  \label{eqn4.44}
\end{align}

\noindent the cuadratic term of Eq. \eqref{eqn1.3}  dominates over
the linear term, and in the computation of the three-point function
Eq. \eqref{eqn1.6}, the loop contributions become dominant. The 
computation of non-Gaussianities in this case goes beyond the
scope of this paper and should be treated carefully elsewhere. (for
discussions on the origin and magnitude of the loop corrections see Refs.
\cite{weinberg05,zaballa06b,byrnes07})

Here we restrict ourselves to use the marginal values allowed for $\fnl$ from
WMAP-III observations and look at the modifications
that large non-Gaussianities bring to the amplitude of
perturbations at the PBH scale.

\begin{figure}[h!bt]
  \begin{center}
    \psfrag{A}{\Large$\log{(\beta_{\rm PBH})}$}
    \psfrag{B}{\Large$\log{\left(\frac{M}{1 \gr}\right)}$}
    \includegraphics[totalheight=.35\textheight]{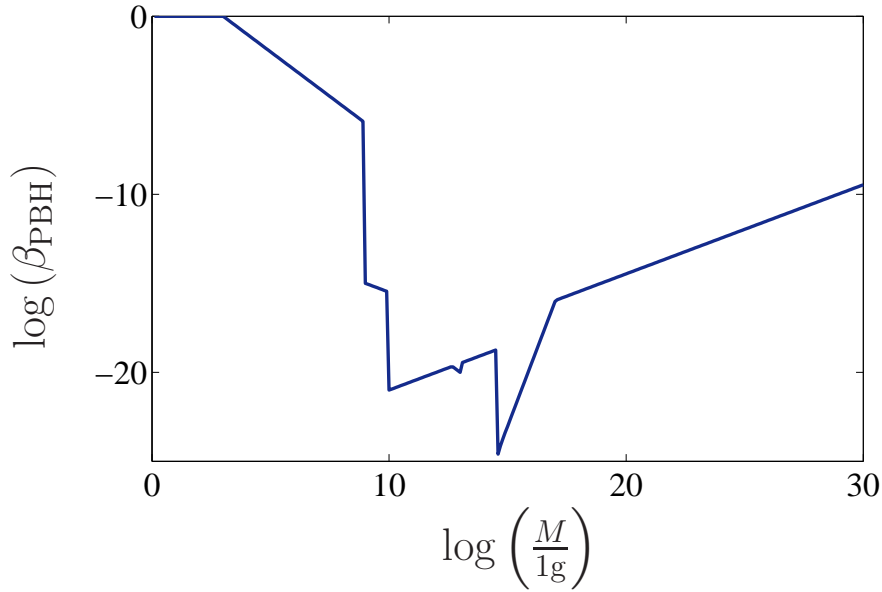}
    \caption{\small{The constrains in Table I are plotted together
	with only the smallest value considered for every
	mass.}}\label{fig2}
  \end{center}
\end{figure}

\begin{figure}[hb]
  \begin{center}
    \psfrag{F}{\Large$\log{(\Sigmar)}$}
    \psfrag{E}{\Large$\log{\left(\frac{M}{1 \gr}\right)}$}
    \includegraphics[totalheight=0.35\textheight]{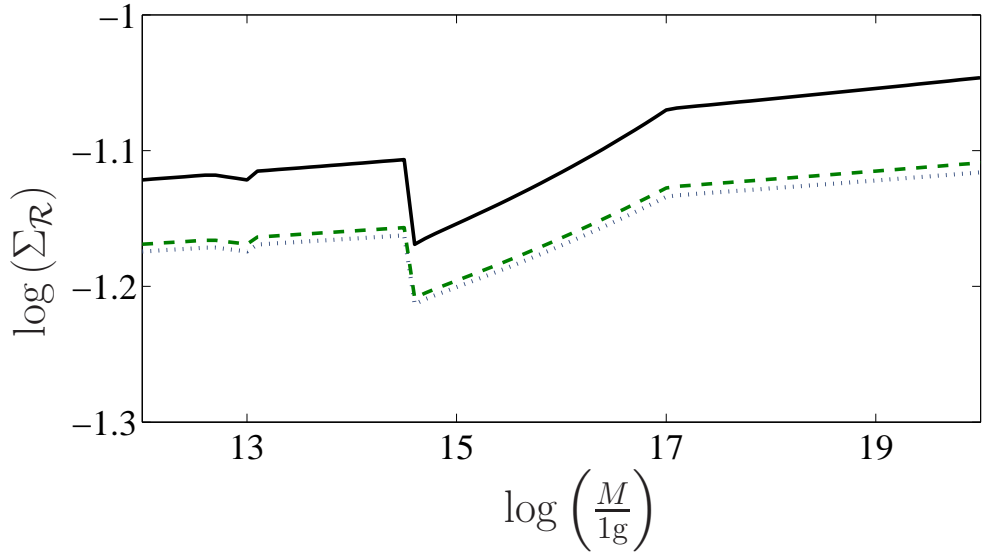}
    \caption{\small A subset of the constraints on $\Sigmar$ from
      overproduction of PBHs 
      is plotted for a Gaussian (black line) and non-Gaussian correspondence
      between $\beta$ and $\Sigmar$, equations \eqref{eqn4.22} 
      and \eqref{eqn4.4} respectively. The green dashed line assumes a
      constant $\fnl = -54$ and the blue dotted line a value $\fnl = - 1
      / \Sigmar^2$.}\label{fig3} 
  \end{center}
\end{figure}

In Fig. \ref{fig2} we plot the set of bounds to the initial mass
fraction of PBHs listed in Table I. The corresponding bounds
on $\Sigmar$ are shown in Fig. \ref{fig3} for the Gaussian and
non-Gaussian cases. Independently of the model of cosmological
perturbations adopted, one can use the observationa limits of $\fnl$ to
modify the bounds for $\Sigmar$ on small wavelengths. The tightest
constraints on $\Sigmar$ comes from perturbations of initial mass $M
\approx 10^{15} \gr$. With the non-Gaussian modification the limit is
$\log{(\Sigmar)} \leq - 1.2$, compared to the Gaussian case
$\log{(\Sigmar)}\leq -1.15$. As shown in Fig. \ref{fig3}, the
modification to $\Sigmar$ cannot be much larger if we use the limit
value of Eq. \eqref{eqn4.44}. 

\section{Final Remarks}\label{sec5}

In this paper we have computed, to lowest order of non-linearity,
the effects of non-Gaussian perturbations on PBHs formation.

We use curvature perturbations with a non-vanishing three-point
correlation to find an explicit form of the non-Gaussian PDF with
a direct contribution from the non-linear parameter $\fnl$.
We have shown how the sign of this parameter determines the
enhancement or suppression of probability for large-amplitude
perturbations. Using the simple slow-roll expression
for $\fnl$ in the context of single field inflation,
We have resolved previous discrepancies in the literature regarding
effects of non-Gaussianity on the abundance of PBHs.

As a second application of the non-Gaussian PDF we have employed the
Press-Schechter formalism of structure formation to determine the
non-Gaussian effects on PBH abundance.  We have shown how the
PBH constraints on the amplitude of perturbations
can be modified when a non-Gaussian distribution is considered.
The maximum variance $\Sigmar$ allowed by PBH constraints is
sensitive to $\fnl$, producing the limit
$\Sigmar(M = 10^{15}\gr) < 6.3 \times 10^{-2}$ for $\fnl = -54$.
This limit is, however, much larger than the observed amplitude at CMB
scales, where  $\Sigmar \approx 4.8\times 10^{-5}$. The order of
magnitude gap between the mean amplitude observed in cosmological
scales and that required for significant PBH formation remains almost
intact and, as a consequence, non-Gaussian perturbations do not modify
significantly the standard picture of formation of PBHs.

\section{ACKNOWLEDGEMENTS}

This work has been financially supported by the
Mexican Council for Science and Technology (CONACYT) Studentship
No. 179026. I would like to thank Prof Bernard Carr for
motivation and major comments to this work. In addition, I would like to
thank Dr David Seery for comments and discussions throughout the time
devoted to this project. I also wish to thank Dr Karim Malik for 
helpful comments on the preparation of the final version of this
paper. Finally I thank Prof David Lyth and the Department of 
Physics at Lancaster University for their hospitality during the
workshop {\it Non-Gaussianity from Inflation}, 5th-9th June 2006.

\newpage
\begin{center}
  {\small
    \text{Table I}
    Constraints on the fraction of PBH density to the density of
    the universe as described in Ref. \cite{lidsey93}
    \begin{align*}
      \begin{array}{l|l|l}
	\hline
	\mathbf{CONSTRAINT \,(\beta \leq)} \,
	& \mathbf{MASS\, RANGE} \,& \mathbf{NATURE} \, \\
	\hline
	\hline
	1.5098 \e{-8} (M / M_{\odot} )^{1/2}
	& 10^{15}-10^{43}\text{g} & \Omega_{PBH}(\text{today})\leq 1\\
	2.1568 \times 10^{-16} (\frac{M}{M_{\odot}})^{1/2}\times
	&\, 3.64 \times 10^{14} -  10^{15} &
	\text{\small X-rays from}\\
	\left\{1 - \left[ 1 - \left( \frac{3.64\e{-9}}{M}\right)^3
          \right]^{1/3} \right\}^{-1} & &
	\text{\small evaporating PBHs}\\
	9.2\e{54} M^{11/2}& 2.51\e{14}-3.64\e{14}
	&  \text{\small evaporated PBHs X-rays}\\
	4.1\e{-3}\left(\frac{M}{\ten{9}\text{g}}\right)^{1/2}
	&\ten{9}-\ten{11}&\text{\small pair production at nucleosynthesis} \\
	6.57\e{-5} \left(\frac{M}{\ten{11}\text{g}}\right)^{7/2}
	&  \ten{11} - \ten{13} & \text{\small Helium-4 Spallation} \\
	4.92\e{-7} \left(\frac{M}{\ten{10}\text{g}}\right)^{3/2}
	&  \ten{10} - \ten{11} & \text{\small Deuterium destruction} \\
	\ten{-15} \left(\frac{M}{1\e{9}}\right)^{-1}
	&  \ten{9}\text{g}-\ten{14}\text{g} & \text{\small CMB distortion} \\
	\ten{-18} \left(\frac{M}{1\e11\text{g}}\right)^{-1}
	&  < \ten{11}\text{g} & \text{\small entropy of the universe}
      \end{array}
      \vspace{0.5cm}
    \end{align*}
  }
\end{center}

\end{document}